\documentclass[pra,twocolumn,nofootinbib,eqsecnum,floatfix]{revtex4}
\usepackage{epsfig}
\usepackage{bm}% bold math
\usepackage{amsmath}
\input epsf
\numberwithin{equation}{section}
\renewcommand{\theequation}{\arabic{section}.\arabic{equation}}

\newcommand{\ltwid}{\mathrel{\raise.3ex\hbox{$<$\kern-.75em\lower1ex\hbox{$\sim$}}}}
\newcommand{\gtwid}{\mathrel{\raise.3ex\hbox{$>$\kern-.75em\lower1ex\hbox{$\sim$}}}}

\begin{document}

\title{Quantum Physics and Human Language\footnote{Dedicated to GianCarlo Ghirardi on his 70$^{\rm th}$ Birthday.}}

\author{James B.~Hartle}

\email{hartle@physics.ucsb.edu}

\affiliation{Department of Physics,\\
 University of California,
 Santa Barbara, CA 93106-9530}

\date{\today}

\begin{abstract}

Human languages employ constructions that tacitly assume specific properties of the 
limited range of phenomena they evolved to describe. These assumed properties are true 
features of that limited context, but may not be general or precise properties of all the 
physical
situations allowed by fundamental physics. In brief, human languages contain `excess
baggage' that must be qualified, discarded, or otherwise reformed to give a clear
account in the context of fundamental physics  of even the everyday phenomena that the
languages evolved to describe. The surest route to clarity is to express the
constructions of human languages in the language of fundamental physical theory, not the
other way around. These ideas are illustrated by an analysis of the verb `to happen'  and the word `reality' in 
special relativity and the  modern quantum mechanics of closed systems.

%\vskip .8in
%\centerline{\Huge  DRAFT 6  \today}
\end{abstract}

%\pacs{PACS }

\maketitle

\begin{quote}
``Mind and world\dots have evolved together and in consequence are something
of a mutual fit.'' \\
\centerline{--- Wm.~James, 1893 \cite{Jam93}}

``We are deceived at every level by our introspection.''\\
\centerline{--- F.H.C.~Crick, 1979 \cite{Cri79}}

\end{quote}

\section{Introduction}

Human languages are features of particular kinds of information gathering and utilizing
systems (IGUSes)\footnote{The term IGUS is broad enough to include both single representatives of biological species that have evolved naturally and certain kinds of mechanical devices. It includes human beings, both individually and collectively as members of groups, cultures, and civilizations. It includes intelligent beings that we might meet in the future. For more discussion see \cite{Gel94}}, living late in the universe, dwelling on a minor planet, circling a
garden-variety star, that is but one of about ten billion stars in a galaxy, that is but one
of about a hundred billion other galaxies within the visible universe.  Human languages
are thus features of our universe  but very special ones. Languages evolved over the
history of our species through a combination of frozen accidents and selection by both
physical and cultural evolutionary pressures. Human languages are adapted to provide
certain highly coarse-grained descriptions of the quasiclassical realm of everyday
experience --- the `world' to which James presumably referred in the above quote. Yet, human languages are not restricted to such quasiclassical descriptions.
Suitably extended, they also permit the discussion of regimes under exploration in contemporary
fundamental physics that are characterized by concepts that can be quite far from those 
native to the quasiclassical realm. This essay explores some aspects of the tension between the domains in which human languages evolved and those to which they can be applied.  In particular, we have in mind  applications to the modern quantum mechanics of closed systems, most generally the universe as a whole. This deals not only with quasiclassical realms but many mutually incompatible ones as well. 

The theses of this essay are these: Human languages employ constructions that tacitly 
assume properties of the limited range of phenomena they evolved to describe. These 
assumed properties are true features of that limited context, but may not be general 
properties of all the physical situations allowed by fundamental physics. In brief, human 
languages contain `excess baggage' that must be qualified, discarded, or otherwise 
reformed to give a clear account in the context of fundamental physics of even the everyday phenomena that the languages evolved to describe. 

It is no more surprising to find that human language contains tacit assumptions than it is to discover that we possess a useless appendix. Indeed these limitations of language  are evidence for its evolution just as our appendix is evidence for human evolution.  Rather, the important circumstance is that the flexible, open-ended, nature of human language allows it to be employed in the discussion of concepts very far from those it evolved to describe with only a few precautions and modifications. This essay is concerned with those precautions and modifications.

The surest route to clarity is to 
express the constructions of human languages in the language of fundamental physical 
theory, not the other way around. Alternatively the constructions can be qualified so that tacit assumptions are made explicit. By `human language'  here we mean roughly the language of everyday discourse. By  `language of physics' we mean roughly the language, terms, concepts, and mathematics found in physics textbooks. Making a precise general distinction will not be necessary. because we will do that explicitly in specific examples. 

There are many different physical theories that could be used to illustrate these
theses, many different linguistic constructions that could be considered, and certainly
many different human languages. For simplicity and clarity we shall focus on limited
examples of each of these possibilities.

We will consider just one human language --- English --- and hope that our analysis of the connection between English and physics extends to other human languages in an
essentially equivalent way. Assuming this equivalence we shall refer just to `human
language'  --- singular rather than plural.

We shall consider just two linguistic constructions. The first is the use of the verb 
`to happen' in its various forms: `happened', `is happening', and  `will happen'. Other existential verbs `to be', `to exist', etc.~may be treated similarly. The other construction is the word `reality'.

We shall consider just two theories: the quantum theory of closed systems \cite{Gri02, Omn94, Gel94} and special
relativity. Clarifying the quantum mechanics of closed systems is the objective of this
essay. But special relativity provides a simple, uncontroversial
example.\footnote{It should be noted that none of these considerations is necessary to
discuss the work of G.~Ghirardi, our dedicatee, as for example in \cite{BC03}.} 

When we refer to `quantum mechanics' or `quantum theory'  in this essay, we mean the decoherent (or consistent) histories quantum mechanics of closed systems, most generally the universe \cite{Gri02,Omn94,Gel94}. This is a modern synthesis of the ideas of many that extends  the work begun by Everett. Decoherent histories quantum theory is logically consistent, consistent with experiment as far as is known, applicable to cosmology, consistent with the rest of modern physics such as special relativity, and generalizable to include quantum gravity (e.g. as in  \cite{Har95c}). In incorporates Copenhagen quantum theory as an approximation appropriate for measurement situations. It may not be the only theory with these properties, but is the most promising of those presently available in the author's opinion.

Briefly the discussion will be as follows:
A conflict arises between special relativity  and the use of `happened', `is happening', and `will happen' in human language. These constructions tacitly assume that there is a division of spacetime into past, present, and future. But special relativity does not provide a unique such division, but rather many of them. The resulting conflict can be resolved by abandoning  `to happen' and using the language of special relativity. Alternatively spacetime can be conventionally divided into past, present, and future and the use of `to happen'  qualified to refer to that convention.

In quantum mechanics, `happened', `is happening', and `will happened' are probabilistic statements to be qualified with the relevant probability, e.g. `happened with probability $p$'.  These probabilities are constructed from those that quantum mechanics assigns to a set of alternative histories of the universe. Human language tacitly assumes that there is a unique such set, mostly the histories of the quasiclassical realm. But quantum theory exhibits  many alternative sets of histories to describe past, present and future which may be incompatible roughly in the sense that position and momentum are incompatible. 
The resulting conflict can  be resolved by  replacing statements which involve `to happen' with statements about quantum mechanical probabilities. Alternatively, `to happen' can be qualified to refer to the particular set of histories which define those probabilities.

This essay is based on the author's experience in explaining and teaching the quantum mechanics of closed systems along with some modest reflection on the subject. It is intended for {\it physicists} as a practical guide to a clearer understanding of this area  by dealing with some of the linguistic tangles that naturally arise. No new physics is considered, only issues concerning the discussion of physics. The ideas expressed are merely the author's opinions on routes to clarity. Others may find different routes work better. This essay is therefore not a review of all that has been written on these questions.  In particular, there is a long history of discussion of related questions in philosophy and linguistics, but this  essay does not pretend to address any of the deep issues that may arise there (see, e.g.\cite{xx}). 

The paper is structured as follows: Section II discusses the use of `to happen' in the
context of special relativity. Section III discusses the use of `to happen' in the quantum mechanical context where there are many mutually
incompatible decoherent sets of coarse-grained alternative histories of the universe.
Section IV ventures to discuss the implications of this for our notions of reality. Section V discusses dispensible words and what to do about them. 
Section VI concludes with practical advice on achieving clarity. 

\section{Spacetime}

Up here, on length scales much greater than the Planck length, the world is
four-dimensional with a classical spacetime geometry. There is neither a unique notion of
space nor a unique notion of time. Rather, from each point in spacetime, there is a family of
timelike directions and three times as many spacelike directions. Spacetime geometry is curved by mass-energy according to the laws of general relativity, but in
sufficiently small patches every geometry is well-approximated by the flat spacetime of 
special relativity.  To simplify our discussion we begin by restricting attention to a 
patch of spacetime where this approximation holds.

First, recall a few basic facts about flat spacetime. Events occur at points. At each
point $Q$ there is a light cone consisting of two parts: The future light cone is the
three-dimensional surface generated by light rays emerging from $Q$. The past light cone
is similarly defined by light rays converging on $Q$. (The labels `past' and `future' are
conventional, but may be conveniently specified by the directions toward the big bang and away from it.) Points inside the light cone of $Q$ are timelike separated from it; points
outside it  are spacelike separated.

The center of mass of a localized IGUS such as ourselves describes a timelike world
line whose points can be labeled by the timelike distance along the curve also known as
proper time.

Consider an event $A$ on your world line and another event $B$ elsewhere. Suppose you are
located along your world line at event $A$. Could you answer the questions: `Will $B$
happen in the future?'; `Is $B$ happening now?'; and `Did $B$ happen in the past?' You
cannot because these questions are meaningless in special relativity without further
qualifications.  The questions presume that spacetime can be divided into past, present
(now), and future just on the basis of the location of $A$ in spacetime. There is no such
division.

More specifically, imagine there are astronauts on Saturn's moon Titan. Questions like `What are the
astronauts doing now?' or `Did they begin breakfast half an hour ago?' are meaningless
without further qualifications. The light travel time between Titan and Earth can be over an hour implying a comparable special relativistic ambiguity in the notion of
simultaneity. There is thus a conflict between familiar constructions of human language
and the facts of special relativity.

The simplest resolution of the conflict is to discard the verbs like, `to happen' and to
formulate physically meaningful questions in the language of physics. Questions like  `Is
$B$ in my future light cone at $A$?', `Is $B$ spacelike separated from me at $A$?', and
`Is $B$ in my past light cone at $A$?' do make sense and have unambiguous answers\footnote{`Is'
here is understood in a tenseless, four-dimensional sense referring to the properties of
spacetime. In a similar way `happen' is sometimes used (mostly by physicists) in a four-dimentional sense of occurring in spacetime. That usage won't be discussed in this essay  but can be a further source of confusion.}.

Another route to resolving the conflict is to retain the verb `to happen'
but to qualify its usage in some conventional way. Specifically, we could divide
spacetime up into a spacelike surface $S$ containing $A$, the future of $S$, and the past
of $S$. Then the questions `Will $B$ happen in the future of $S$?', `Is $B$ happening on
$S$?', and `Did $B$ happen in the past of $S$?' do make sense and have definite answers
depending on where $B$ is in spacetime. They are made meaningful by the qualifications
referring to the spacelike surface $S$. However, there are an arbitrarily large number of
spacelike surfaces containing $A$ and the answers to the questions can be different for other spacelike surface $S^\prime$ when  $B$ is spacelike separated from $A$.
Unqualified constructions involving `to happen' are meaningless, when qualified they 
have many different meanings\footnote{Using the language of physics and employing appropriate qualifications are not to only ways of resolving the conflict between special relativity and human language. Raphael Sorkin tells the author that there have been various proposals for {\it reforming} human language. These include redefining `now' to mean `spacelike' and using adverbs rather than verb forms to indicate tense, as in `It happen spacelike.' The author guesses that the two resolutions discussed in the text have better chances of success. }.

Introducing a coordinate system in our patch of spacetime is another arbitrary way of
fixing a convention to give `to happen' a qualified meaning. Riemann normal coordinates
based at the point $A$ are a simple example if they cover the whole patch. If the
geometry of the patch happens to be exactly flat, these reduce to a choice of a
particular Lorentz frame. Past, present, and future are now defined by the timelike
coordinate (call it $t$) being greater, equal, or less than its value at $A$. As a bonus, a 
coordinate system provides a convention for answering further questions of the form `Does
$B$ occur before $C$ or afterward?' which would be ambiguous if unqualified when $B$ and
$C$ are spacelike separated.

The satellites comprising the Global Positioning System (GPS) are an example of a
collective IGUS that effectively employs a coordinate system to define temporal
relations. Both the special relativistic effects arising from the velocities of the
satellites, and the general relativistic effects arising from the slight curvature of
spacetime in the vicinity of the Earth are important for GPS operation \cite{Ash02}. The
system would fail in about an hour if these were not both accounted for. Precise
agreement among the satellites on a notion of simultaneity is needed. To define that the GPS uses a version of a standard set of coordinates for the weak field metric of general relativity, centered in the Earth, with spatial axes pointing towards fixed stars,  and the  time coordinate normalized so that that  on the Earth's geoid  (approximately the ocean surface) it coincides with the time of clocks co-rotating with the Earth  there.  It is unlikely that the satellites are
employing the  verb `to happen' in any communication. But, if they did, they could
define it by the time of their effective coordinate system.

Human IGUSes were using constructions like `will happen', `is happening', and `did happen'
long before they had accurate clocks, and certainly before there was a precise notion of
coordinate system. How would the notion of `present' or `now' that is implicit in these
constructions be described in the language of physics? The author has discussed the physics
of `now' in \cite{Har05}. The next few paragraphs summarize some of these ideas.

Human IGUSes have an individual notion of `now' which can be modeled as a feature of their conscious focus on their most recently acquired information. This is already an approximate idea only defined up to the time scale of human perception\footnote{This is of order .1s   e.g \cite{Wesxx}.} which we denote by $\tau_*$. Individual IGUSes can agree on `what is happening now' by
reporting their current observations and checking the reports they receive against their
individual notions of `now'. The result is an approximate, imprecise, common present useful
in everyday circumstances.

Agreement on a common present can be reached by a group of IGUSes in a patch where spacetime is approximately flat if the 
following contingencies are met:
\begin{enumerate}

\item The relative velocities of the IGUSes are small compared to the velocity of light.

\item The light travel time between IGUSes in a Lorentz frame in which they are nearly at
rest is small compared to the time scale $\tau_*$ characterizing perception.

\item The time scale for perception $\tau_*$ is short compared to the time scales on which
interesting features of the IGUSes' environment vary.

\end{enumerate}
Contingency (2), based on (1), means special relativistic ambiguities in the meaning of
simultaneity are negligible in the construction of a common present. Contingency (3) means
that the ambiguity in the `now' of each IGUS is negligible in this construction.
Condition (1) means that agreement can be reached over an interesting length of time. 

Under these contingencies, a collection of IGUSes can construct a common present, but it is
a present that is local, approximate, and contingent on their relation to each other and to
their environment. All the contingencies are easily satisfied for human IGUSes on Earth.
However, astronauts on Titan will not be able to participate in this common `now';
contingency (2) will be violated. Further conventions will be needed to use the verb 
`to happen' in talking to them.

The purpose of this section was not to analyze human linguistic constructions in any detail.
Rather, it was to emphasize that certain of these constructions referring to time are
ambiguous even in the context of our understanding of the physics of spacetime through special
and general relativity. The existence of such ambiguities is not surprising given that the
language evolved to describe limited and specific circumstances in which conditions (1)- (3) hold. The language
implicitly assumes certain features of these limited circumstances summarized roughly by
the three contingencies above. The ambiguities can be
resolved either by replacing parts of the human language with the language of physics, or by using that
language to specify conventions that resolve the ambiguity.

As our understanding of spacetime progresses to ever more general contexts, the ambiguities
in the use of human language can be expected to become larger. In a quantum theory of
gravity, there is no fixed spacetime geometry. Rather, geometry is a quantum variable,
fluctuating and without definite value.  Then, even statements like `This happened in my
past light cone' become meaningless without further qualification. Beyond that, some
explorers expect that spacetime will only be an coarse-grained phenomenon of some
deeper level of description \cite{Sei06}. How will `to happen' be unambiguously defined then?

\section{The Quantum Mechanics of Closed Systems}

\subsection{A Model Universe}

To keep the discussion manageable, we consider a closed quantum system, most generally, the
universe, in the approximation that gross quantum fluctuations in the geometry of spacetime
can be neglected. The closed system can then be thought of as a large (say $\gtwid$ 20,000
Mpc), perhaps expanding, box of quantum fields moving in a fixed background spacetime.
Everything is contained within the box, in particular galaxies, planets, observers, and
observed (if any), measured subsystems, and the apparatus that measures them. This is a model of the
most general physical context for prediction.

The fixed background spacetime means that the notions of time are fixed and that the usual
apparatus of Hilbert space, states, and operators can be employed in a quantum description
of the box. The essential theoretical inputs to the process of prediction are the
Hamiltonian $H$ governing evolution and the initial quantum state $|\Psi\rangle$. These are
assumed to be fixed and given by fundamental theory. We assume that $H$, $|\Psi\rangle$, and
the operators representing alternatives can be described in terms of a set of fundamental
fields and their conjugate momenta.  For definiteness, we work in a fixed Lorentz frame
whose time is $t$.

All the special relativistic concerns regarding  the use of `to happen' discussed in
Section II arise in this context,
but we resolve these by fixing the Lorentz frame in order to concentrate on 
quantum mechanical issues.

For the reader not familiar with it, a simplified,  bare-bones account of the quantum mechanies
of closed systems is given in Appendix A. However, very little of even the modest detail
given there is necessary for the present discussion. The points essential for the present discussion are these: 

\begin{enumerate}
\item  The assumed inputs are theories of the quantum dynamics ($H$)
and the quantum initial state  $(|\Psi\rangle)$.
\item The outputs are the probabilities of the individual members of sets of
alternative coarse-grained histories of the closed system. Consistent probabilities are predicted only  for sets of histories for which there is  negligible quantum interference between all pairs of histories in the set as a consequence of $H$ and $|\Psi\rangle$. Such sets of histories are said to {\it decohere} and  are called {\it realms} for short. 
\item There are no non-trivial completely fine-grained realms. Coarse graining is therefore necessary for decoherence. Some realms are {\it compatible} in the sense that they can be related by the operations of fine and coarse graining. But quantum theory also exhibits {\it incompatible} realms.  Two realms are mutually {\it incompatible} if there is no common finer-grained realm of which they are both coarse
grainings. Realms defined by coarse grainings of incompatible variables such as position and momentum provide simple examples. 
\item {\it Quasiclassical realms\footnote{\rm We use the term {\it quasi}classical realm to emphasize that the classical behavior is probabilistic and on occasion significantly interrupted by quantum mechanics.}} exhibit the regularities of classical physics and in particular the approximate correlations in time summarized by effective classical equations of motion. At a sufficiently fine-grained level, quasiclassical realms are defined by coarse grainings of familiar {\it quasiclassical variables} such as averages of energy, momentum, an number over suitable volumes. The quasiclassical realms of everyday experience are a subset of the
totality of realms provided by quantum theory for a description of the universe.
\item Quantum theory does not distinguish between its different realms, although
IGUSes may distinguish between them by their utility, as for example in the almost  exclusive focus of human IGUSes on quasiclassical realms.
\end{enumerate}
If any of this is not immediately clear, the reader should consult  Appendix A.

\subsection{Probabilities}

Probabilities are measures of ignorance in classical physics, but in quantum physics they
are fundamental. This subsection is devoted to reconciling human language which
incorporates assumptions of classical certainty with a probabilistic fundamental theory.
To this end, we restrict the discussion of this subsection to one realm. We begin by
reviewing a little notation explained more fully in Appendix A.

We consider histories that are sequences alternatives labeled by $\alpha_1, \alpha_2,
\cdots, \alpha_n$ at a series of times $t_1, t_2, \cdots, t_n$. The probabilities of
these are given by
\begin{equation}
p(\alpha_n, \cdots, \alpha_1) = \parallel P^n_{\alpha_n} (t_n) \cdots P^1_{\alpha_1}
(t_1)\, |\Psi\rangle\parallel^2
\label{threeone}
\end{equation}
where the $P^k_{\alpha_k}(t_k)$ are Heisenberg picture projections onto these
alternatives. The label $k$ denotes the particular exhaustive {\it set} of exclusive
alternatives, $\alpha_k$ labels the particular alternative within the set, and $t_k$ 
is the time. To
conserve on notation, we denote individual histories by $\alpha\equiv (\alpha_1, \alpha_2,
\cdots, \alpha_n)$ and denote the corresponding chains of projections by $C_\alpha$ so that 
\begin{equation}
p(\alpha)=\parallel C_\alpha |\Psi\rangle\parallel^2
\label{threetwo}
\end{equation}
is a shorthand for \eqref{threeone}.

Conditional probabilities for alternatives $\alpha$ given another alternative $\beta$ are constructed in the usual way
\begin{equation}
p(\alpha|\beta) \equiv p(\alpha, \beta)/p(\beta)\, .
\label{threethree}
\end{equation}
Most useful probabilities are conditional. Suppose, for instance, we know certain data
$d$ about the universe at a present time $t_0$ represented by a projection $P_d(t_0)$. The
predictions for future histories of the universe given this data are specified by the
conditional probabilities 
%$\alpha_{\rm fut}\beta_{\rm pst}$
\begin{equation}
p(\alpha_{\rm fut}|d)=\frac{\parallel C_{\alpha_{\rm fut}}
P_d(t_0)|\Psi\rangle\parallel^2}{\parallel P_d(t_0)|\Psi\rangle\parallel^2}
\label{threefour}
\end{equation}
where the $\{C_{\alpha_{\rm fut}}\}$ represent an exhausitive set of alternative
histories to the future of $t_0$. Similarly, the probabilities for retrodiction of the 
past are given by
\begin{equation}
p(\beta_{\rm pst}|d)= \frac{\parallel P_d(t_0)\, C_{\beta_{\rm
pst}}|\Psi\rangle\parallel^2}{\parallel P_d(t_0) |\Psi\rangle\parallel^2}
\label{threefive}
\end{equation}
where $\{C_{\beta_{\rm pst}}\}$ represent an exhaustive set of alternative histories to the
past of $t_0$.  In each case, we assume that the set of histories consisting of $P_d$ and
the $C$'s is decoherent.

The alternatives $P_d(t_0)$ can refer to the data possessed by an  IGUS at a
time $t_0$ along its own history. Similarly, $\alpha_{\rm fut}$ and $\beta_{\rm pst}$
can refer to data possessed by that IGUS in the future or past of $t_0$. Thus, it is possible
to provide probabilities that answer questions like ``Given that I observe a tree here today,
what is the probability that there was a tree here yesterday?'' or ``Given that I
observe a tree here today, what is the probability that I will observe  a tree here
tomorrow?'', or even ``Given that I observe a tree here today, what is the probability
that there is a tree here today?'' (This the probability that our observations
don't deceive us). 

Like all assertions in a probabilistic theory, those involving he verb `to happen' should always be qualified by a probability. For instance, from present data that includes texts giving 55 BC
as the date of the Roman invasion of Britain we would like to infer that Caesar did
invade Britain in 55 BC. But the probability for this is not unity. There is some
probability that the texts are forgeries, or contain propagating mistakes, or that the
ink on their pages made a quantum transition from a configuration spelling a different
date. If the probability is sufficiently close to unity we say simply that the Roman
invasion of Britain {\it happened} in 55 BC. Similar qualifications are needed for `is
happening' and `will happen'. 

This need for qualifying `happen'  because of probabilities is a trivial observation. Such qualifications are
generally needed
in classical physics as well. Probabilities are inescapable as a practical matter because of ignorance  of present data or inability to determine
classical evolution. Quantum fluctuations add one more source of uncertainty which
is often negligible in everyday circumstances, as in probability for the date of the Roman invasion of Britain. In the next subsection we consider a more serious need for qualification. 

\subsection{Incompatible Realms}

Human IGUSes focus almost exclusively on coarse-grainings of the quasiclassical realm.
Our senses are adapted to perceive quasiclassical variables and our language is adapted
to describe quasiclassical histories\footnote{Explanation for our quasiclassical focus
can be sought in the physical structure of human IGUSes and the origin of that structure.}.
%to coarse-grainings of the quasiclassical realm is plausible because such realms exhibit
%enough regularity over time to permit the generation of models (schemata) with
%significant predictive power \cite{GH90a}.}

As we have already mentioned, the quantum universe exhibits distinct realms which are
incompatible in the sense that there is no finer-grained realm of which they are
coarse-grainings. Questions, answers, predictions, retrodictions, etc.~are
all in the context of a particular realm which must be specified to understand what they
mean.

Consider by way of example the reconstruction of a past history of the universe from data we have gathered in the present together with the theory of the initial state\footnote{For more on the reconstruction of the past in quantum mechanics, see
\cite{Har98b}.}. As mentioned above, that is accomplished by retrodicting the probabilities of past events from the given present data using the conditional probabilities \eqref{threefive}.  

Suppose present data is given in  quasiclassical variables.
The most familiar and useful retrodictions from this data  are made using quasiclassical past histories.
With these we retrodict the date 55 BC for the first Roman invasion of Britain from
present textual records. We use present observations of the planets to reconstruct their
past trajectories. We use fossil
records to estimate that there is a high probability that dinosaurs roamed the Earth 150
million years ago. We infer that matter and radiation were in thermal equilibrium at the
beginning of the universe from the present values of the Hubble constant, spatial
curvature, and the temperature of the cosmic background radiation. These are all past histories that are members of quasiclassical realms based on coarse-grainings of  quasiclassical variables.

But in quantum theory there is no unique past conditioned on given present data.  Incompatible past realms can provide
different stories of what happened. A striking, if artificial, example of this is
provided by the three-box model introduced by Aharonov and Vaidman for a different purpose \cite{AV91}.

Consider a particle that can be in one of three boxes, $A$, $B$, $C$ in corresponding
orthogonal states $|A\rangle$, $|B\rangle$, and $|C\rangle$. For simplicity, take the
Hamiltonian to be zero, and suppose the system to initially be in the state
\begin{equation}
|\Psi\rangle\equiv \frac{1}{\sqrt{3}}\ (|A\rangle + |B\rangle + |C\rangle)\, .
\label{threesix}
\end{equation}
Suppose for present data at $t_0$ the particle is in the state
\begin{equation}
|\Phi\rangle \equiv \frac{1}{\sqrt{3}}\ (|A\rangle + |B\rangle - |C\rangle)\, .
\label{threeseven}
\end{equation}
Denote the projection operators on $|\Phi\rangle$, $|A\rangle$, $|B\rangle$, $|C\rangle$
by $P_\Phi$, $P_A$, $P_B$, $P_C$ respectively. Denote by $\bar A$ the negation of $A$
(``not in box $A$'') represented by the projection $P_{\bar A}=I-P_A$. The negations
$\bar \Phi$, $\bar B$, $\bar C$ and their projections $P_{\bar\Phi}$, $P_{\bar B}$, and $P_{\bar C}$
are similarly defined.

From the present data $|\Phi\rangle$ and the initial condition $|\Psi\rangle$ let us ask
whether the particle was in the box $A$ at a time earlier than $t_0$. (The exact values
of the times are unimportant since $H=0$. Only the order matters.) The relevant past realm
consists of the histories
\begin{equation}
P_\Phi P_A\, ,\ P_\Phi P_{\bar A}\, ,\ P_{\bar\Phi} P_A\, ,\ P_{\bar\Phi} P_{\bar A}\, ,
\label{threeeight}
\end{equation}
and is easily checked to decohere exactly. The conditional probabilities for $A$ and
$\bar A$ given $\Phi$ can be calculated from \eqref{threefive} with $P_d=P_\Phi$ and
$\{C_\alpha\}=\{P_A\, ,\, P_{\bar A}\}$. The result is
\begin{equation}
p(A|\Phi)=1\ , \quad p(\bar A|\Phi)=0\, .
\label{threenine}
\end{equation}
Thus, we can say  {\it  in this past realm} that  the event that the particle was in box $A$ in the past happened. 

But an examination of \eqref{threesix} and \eqref{threeseven} shows that both initial
condition and present data are symmetric under interchange of $A$ and $B$. Therefore,
using the decoherent set of histories
\begin{equation}
P_\Phi P_B\, ,\ P_\Phi P_{\bar B}\, ,\ P_{\bar \Phi} P_B\, ,\ P_{\bar\Phi} P_{\bar B}
\label{threeten}
\end{equation}
we can compute
\begin{equation}
p(B|\Phi)=1\ , \ p(\bar B|\Phi)=0\, .
\label{threeeleven}
\end{equation}
Thus, we can say {\it in this past realm} that the event that the particle was box $B$ happened.

There is no contradiction because the sets of histories \eqref{threeeight} and
\eqref{threeten} are incompatible realms. The finer-grained set of histories describing
both $A$ and $B$ is
\begin{equation}
P_\Phi P_A P_B\, , \ P_\Phi P_A P_{\bar B}\, ,\ P_\Phi P_{\bar A} P_B\, , \cdots ,
\ \text{etc.}
\label{threetwelve}
\end{equation}
But this set does not decohere. The inference ``if in $A$ then not in $B$'' cannot be
drawn since there are no consistent probabilities for it.

Eqs.~\eqref{threeeight} and \eqref{threeten} do not exhaust the possible realms defining
possible pasts for the present data $P_\Phi$. For example, we could consider
\begin{equation}
P_\Phi P_\Psi\, ,\ P_\Phi P_{\bar \Psi}\, ,\ P_{\bar\Phi} P_\Psi\, ,\ P_{\bar \Phi}
P_{\bar \Psi}
\label{threethirteen}
\end{equation}
This is trivially decoherent with easily anticipated probabilities, but also clearly
distinct from \eqref{threeeight} and \eqref{threeten}. In this past realm we could say that
$P_\Psi$ happened rather than anything about the above alternatives.

The usual use of `happened' assumes that there is only one realm. In a theory that
permits incompatible pasts, its use must be reformed. As with the other conflicts between
human language and fundamental physics discussed in this paper, there are two routes to
improving precision and clarity. One is to use the language of physics and speak of the
past in terms of the conditional probabilities in different past 
realms. The other route is to qualify `happen' so it refers to a particular realm. For
instance, in the three-box example we could say that `the event that the particle was in
box $A$ happened in the realm that referred to $A$' (or whatever other
characterization of \eqref{threeeight} one prefers) and similarly for $B$.

If `happened' means high probability for an event in the past  conditioned on certain present data, the
above examples show that different events can have happened in different incompatible
pasts, even seemingly contradictory events.  If someone asks you `What happened yesterday?' you should strictly speaking respond with the question `In what realm?'. 

It should be stressed, however, that the
same event cannot have happened in one realm and not  have happened in another.  The
probability for a past history is given uniquely by \eqref{threefive} in all the realms
of which it is a member. If it is high in one realm, it is high in all the others. In
this sense `happened' is non-contextual.

Needless to say, similar considerations apply to `happening' and `will happen'.  We next turn to the implications of all this for the word `reality'.

\section{Reality}

The words `real' and  `reality' are used in many different ways in human language. In the following we attempt to draw crude distinctions  between a few of these without suggesting that other distinctions are not possible. The general point is that notions of reality reside in the models (schemata) that IGUSes construct of the world around them, both individually and collectively. Different models have different notions of reality. Therefore, when using the words `real'  or `reality' it is important for clarity to specify which model is being referred to.

\subsection{Everyday Physical Reality}

Everyday notions of physical reality arise from the agreement among human IGUSes, both individually and collectively, on their observations and on the models of the world
(schemata) that they infer from them. 
These models are formed from the gathered data by processes of selection, communication, and
schematization, consistent with built-in biases. The models are constantly updated as the IGUSes
acquire new information, integrate it with previous experience, infer new useful regularities, and check the model against other schemata.  The everyday notions
of physical reality reside in these models.  This is the reality  of tables and chairs, stars and galaxies, biological species, fellow humans,  and the records of experiments revealing  quantum phenomena, among many other things. These are the notions of reality  that human language evolved to describe and assist in constructing. Explaining the regularities found in  such models is an important objective of science.  
The limits of physical reality are illustrated by the lack of agreement on mirages and
illusions, and by the delusions of schizophrenics who are said to be `out of touch with
reality'. In human language, this everyday physical reality is often what is meant when the word isused without qualification.

How do we understand the agreement among human IGUSes on the facts of their physical
reality in a quantum universe characterized fundamentally by the distributed probabilities
of the alternative histories of a vast number of incompatible realms? The simplest
explanation is that human IGUSes are all making observations utilizing coarse-grainings of the quasiclassical realms in order  to exploit the quasiclassical
regularities that these realms exhibit. They thus are adapted to develop schemata in more or less
the same way. Occasionally they slip up as when they are subject to delusions, or see
canals on Mars, or find ghosts under the bed. But they agree, by and large.  Indeed, we could
not function in social units without this agreement; the ability to construct a common physical reality must have been a highly adaptive trait. Plausibly, many other IGUSes on Earth, such as dolphins and ants,  make use of similar coarse grainings. 

If we find intelligent life on other planets, will they have the same notion of physical
reality that we do? It's plausible that many kinds of IGUSes will have evolved to exploit
the regularities of quasiclassical realms as we have.  In that case, we can expect
to reach agreement with them. But could there be IGUSes focused on coarse-grainings of a
distinct, incompatible realm with a correspondingly different notion of physical reality? The statistics of the schemata of extra-terrestrial IGUSes will constitute a test of the conjecture of the adaptive utility of the quasiclassical realms. 

Closer to home, we imagine we could construct mechanical IGUSes (robots) that have notions of reality  differing from the human kind. Even restricting to input data streams that are coarse grainings of a quasiclassical realm, we imagine that we could construct robots that create different schemata from that data. For instance the built in biases for selecting what to schematize and the rules for how to schematize could both be varied. A thermostat is a very simple example of an IGUS with a restricted schema. It should be possible to construct IGUses (robots) with schemata that do not employ our past, present future organization of temporal information\footnote{Our conscious focus on the most recently acquired data is plausibly the reason that the present is sometimes characterized as more `real' that the imperfectly recorded past and the unknown future. What is meant presumably is that present data figures more prominently and accurately in robot's schema. However, the `NS' robot of \cite{Har05} would  treat the past and present equally, and the `AB' robot would have some premonitions of the future.}\cite{Har05}. Science fiction abounds with robots that construct different schemata. It is also not beyond possibilty that we could construct robots utilizing  non-quasiclassical input data streams. Such robots would consequently have qualitatively  different schemata and qualitatively different notions of everyday reality. 

\subsection{The Realities of Physical Theories}

As Bohr said, ``the task of science is both to extend the range our experience and reduce it to order'' \cite{Boh61}.
Elementary notions of physical reality are extended  by fundamental physical
theories.  These are realities that are agreed to by physicists and reside in physics
literature.  Indeed, there is very little, if any, distinction between the model itself
and the notion of reality which follows from it\footnote {That is possibly the reason that `reality' is so little discussed in physics textbooks --- it is already implicit in the model under discussion.}. 

 We hope that the reality of our fundamental theoretical frameworks are objective because they summarize the universal regularities of the universe independently of any selection by us. In particular we hope that they would be agreed to by other IGUSes we might meet that are interested in physics whether they share our notion of everyday physical reality or not. 

The everyday physical reality described in Section A is an approximate, particular feature of these fundamental models arising from particular coarse grainings and particular initial conditions. Characterizing the emergence of physical reality already raises interesting questions  in classical physics and more profound ones in quantum mechanics.

A fundamental classical model for the universe would consist of the fine-grained histories of particles and fields evolving from an initial condition according to deterministic laws. These fine-grained histories can be specified by giving the coordinates and momenta of both particles and fields as a function of time\footnote{More precisely they are specfied by giving these values on a foliating family of spacelike surfaces in a curved spacetime obeying the Einstein equation when general relativity is taken into account.}. An initial condition could be specfied by giving a distribution function for these variables at an initial time. The important point for present considerations is that the reality of classical theory consists of a unique family of fine-grained histories. 

The constituents of the everyday notion of physical reality are not atoms, molecules, and electromagnetic fields, but rather tables and chairs, stars and galaxies, etc. more generally the forms, velocities, and locations of individual objects. Indeed such notions were used by our species long before atoms, molecules, and fields were discovered. Rather, everyday physical reality arises from the fundamental classical model by appropriate coarse-graining. For instance, coarse-graining by quasiclassical variables such as  the averages over suitable volumes of densities of approximately conserved quantities such as energy, momentum, and number leads to phenomenological equations of motion such as the Navier-Stokes equation for a wide class of initial conditions (see, e.g. \cite{Fos75}). Further coarse-graining is needed to define individual physical objects such as particular trees. The important point is that everyday physical reality is an approximate  feature of classical physics contingent on particular choices of coarse-graining. 

The story of the emergence of everyday physical reality from the quasiclassical realms in quantum theory is similar to that in classical physics. The coarse-grainings defining the relevant sets of histories are the same. The two new features are the following: First the initial quantum state must be such that the set of coarse-grained histories decoheres. Second, the correlations in time that define classical determinism are now only approximate --- occuring with high probability in particular initial states\footnote{For some further discussion and models see e.g. \cite{GH93a,BH99,Hal98,Hal03}.}.

What is very different from classical physics is the reality of the fundamental quantum theory. There are no non-trivial fine-grained decoherent sets of histories for quantum theory as there are for classical physics \cite{GH06}. Coarse-graining is necessary for decoherence, and there are many different coarse-grainings not all of which are compatible. As we have already discussed,  quantum theory therefore provides many mutually incompatible realms of which the quasiclassical ones are a small subset. The theory does not distinguish between these, although IGUSes may do so. In the three-box example in Section III.C,  the past realm that refers to box $A$ is no more or less `real' than the incompatible past realm that refers to box $B$. The reality of quantum theory may rather be said to consist of all the different possible realms\footnote{The far reaching consequences of this generalization of  the reality of classical theory have been stressed in \cite{DK96}}. 

To use the word `reality' in the context of quantum theory  without qualification is to risk confusing everyday physical reality which is constructed from quasiclassical realms with the reality of quantum theory which consists of all realms.

\subsection{Other realities}
Human IGUSes exhibit a wide range of individual notions of  reality that go beyond physical reality by  incorporating in their schemata  the supernatural, prejudice, revelation,  hearsay, wishful thinking, and the like along with scientific evidence.  Today
one can find human IGUSes of different persuasions on the reality of UFO's, paranormal
phenomena, and biological evolution. Beyond these individual notions, there are many
other kinds of agreed-upon models constructed by human IGUSes, each defining a notion of
reality. For instance, a notion of mathematical reality arises from the agreement
of mathematicians on axioms which, as G\"odel put it \cite{Godxx}, `force themselves on us as
being true'\footnote{When we encounter other intelligent beings we can confidently predict that
they will have the same arithmetic that we do.  But will they have ZFC?}  

\subsection{Usage} 
The above discussion probably does not exhaust the uses of the word `reality' in human language. But it does illustrate that its meanings are diverse. What they seem to have in common is an agreement by at least some IGUSes on a model for some class of coarse-grained physical phenomena. This diversity of meanings can lead to confusion if the word is used without qualification. In particular it is important to distinguish between everyday physical reality and the notions of reality provided fundamental physical theory.
That is especially the case if everyday experience is  special and
contingent among many other possibilities as it is in quantum theory. The trend in fundamental physics today seems to lie in the direction of increasing disparity between everyday physical reality  and the reality of the fundamental theory . If that trend continues, maintaining appropriate linguistic qualifications will become even more necessary for clarity. 

\section{Dispensible Words}

Dispensible words are ones that can be added to or deleted from an exposition of a physical
theory without effecting the theory's experimentally verifiable predictions or its utility. 
Such words can be important
for motivation, for evoking analogies, for building intuition, and for suggesting future
research. In short, they can help understanding. But they can also be confusing, the
source of false problems, and an obstacle to understanding. Not surprisingly, some
dispensible words arise from natural constructions in human language. Also, not
surprisingly, many of the confusing dispensible words occur in quantum
theory.\footnote{For one reason this might be so, see \cite{Har05a}.} 

It is useful to know when language describing a theory is dispensible and when it is not. In particular, a question of whether or not dispensible words are appropriate will not be settled by experiment. 

There is a simple test for dispensible words: Dispense with them and see if the
predictions of the theory are unchanged. The author has usually found that dispensing with
the dispensibles is a route to clarity. This section
illustrates these ideas with two examples from quantum mechanics. 

\subsection{Probabilities `to happen'}

Consider the two sentences, `The probability of rain this afternoon is 80\%.' and `The
probability for rain to happen this afternoon is 80\%'. To the author these two sentences
mean the same thing in any understanding of probability of which
he is aware. His decision whether or not to carry an umbrella would be the
same under either assertion. The words `to happen' are dispensible.

%In discussions of the quantum mechanics of closed systems, the author is sometimes asked
%whether we could say that one history happens in each possible realm? By
%by applying `happen' to a history, the questioner is using the word in a four-dimensional
%sense and not in the temporal sense discussed elsewhere in this essay. The author's answer is `yes', because the making the assertion or not does not affect the experimental  consequences of the theory.  To employ the phrase risks confusion,
%however, unless some understanding is reached that `happen' must be qualified by the realm to which it refers along the lines discussed in Section III.C. 
%Some might assert that all
%the histories in a given realm happen (see below). That, however, is more confusing
%because it does not reduce to the ordinary meaning of the word when discussion is
%restricted to a quasiclassical realm.

Qualifying `the probability of $A$' so that it becomes `the probability of $A$ to happen' can  be a source of confusion becaue it suggests to some that `happening' is a physical process. Indeed, some have held that quantum theory is incomplete 
until it explains why, in a set of alternatives for which it supplies probabilities, one of them
`happens' (or is observed).  Quantum theory as developed here has no such mechanism and yet is consistent with all experiment, as far as we know, through the probabilities which are its only output.

\subsection{Equally real histories}

Another candidate for dispensible words is the statement, `all the histories in a given 
realm are equally real'.\footnote{This is a translation of the statement `all the other
worlds are equally real' where what is meant by world is what is called history here even if it consists only of an alternative at one time.} This statement has a simple meaning if `real' is understood to refer to the theoretical model --- the reality of the theory as discussed in Section IV.C. In that context the statement could be rephrased as `quantum mechanics does not distinguish between the
histories in a given realm except by their probabilities.'  Indeed, using the word `real' in this sense, it would be possible to correctly say that `all the realms of quantum theory are equally real'.  This could be similarly rephrased as `quantum theory does not distinguish between different realms'. In each case, the first statement can be replaced by the second without affecting the predictions of the theory for experiment.  The second formulations are easier to understand for many physicists. But all these statements are dispensible as the exposition of quantum theory given in this article shows.
 
Deutsch  \cite{Deu97} especially has stressed the naturalness and interpretative value of the equal status of histories in a given realm, for instance, for understanding the power of quantum computation. Such advantages, however great, do not alter that fact that words like `all the other histories are equally real'  can be dispensed with without affecting the  experimental implications of the theory.

As discussed in Section IV, the linguistic difficulty with using the word `real' is that
it can mean different things in different circumstances even in physics. Maintaining that `all the histories in a set are equally real' risks confusion unless the meaning of `real' is explained. The author's experience is that otherwise it can be confused with the everyday physical reality discussed in Section IV.A . The resulting conflict can be an
obstacle for some to accepting Everett's powerful and natural idea of taking the state of
the universe seriously.

\section{Advice}

The aim of fundamental physics is to find the laws governing the reqularities exhibited universally  by {\it all} physical  systems, without exception, without qualification, and without approximation. The search for these laws has  been seriously underway at least since the time of Newton. Classical mechanics, Newtonian gravity, Maxwell's electrodynamics, the atomic theory of matter, special and general relativity, quantum field theory, superstring theory, the quantum theory  of  geometry,  and quantum cosmology are just some of the milestones in the history of this search. As new regimes of experiment and observation have been explored, more general theoretical frameworks have evolved. Old theories have become effective theories applicable in limited circumstances of the extended context. 

Excess theoretical baggage is typically shed in this process of generalization \cite{excess}. Ideas that were once accepted as fundamental, general, and inescapable have come to be seen as consequent, special, and dispensible. Examples from the history of physics are an Earth centered cosmology, a single universal time, an exact second law of thermodynamics,  fixed Euclidean spatial geometry, and a quantum mechanics restricted to measurements. These ideas were not true general features of the world, but only perceived to be general because of our special universe, our special place in it, and the limited range of our experience. 

Over this history, the models provided by fundamental physical theory have moved far beyond our everyday notions of physical reality.\footnote{ Excess baggage in the language of physics is also typically shed in the process of extension and generalization of physical theory. Indeed, the relationship between the languages of two effective theories often presents problems similar to those between the human language and the language of physics that we have discussed. For example, applying the language of classical physics to quantum mechanical situations can lead to paradoxes which can be resolved by sticking to the language of quantum theory \cite{Aha05,Gri02}. By and large, however these physics linguistic problems have not caused much difficulty and, in any event, are not the subject of this essay.}
The evolution of physical theory beyond everyday notions of physical reality  has complicated its description using a human language that is adapted to that everyday reality. That has been the subject of this essay.

By way of conclusion the author offers a few words of advice on routes to clarity in the face of the disparity between the languages of fundamental physics and that of human IGUSes. This advice is not directed to  how to {\it find} a fundamental theory. Rather is is only about how to deal  with the conflicts with human language that may arise.

\begin{itemize}

\item {\it Identify Theoretical Excess Baggage.}  Remember that ideas that were fundamental and obvious in one theory can become emergent and dispensible in a more general one. The idea that there is a unique past, present, and future defined by physics is one example that we have discussed. Another is the idea that there should be a unique theoretical  reality specified by one decoherent set of alternative fine-grained histories  rather than many different ones that may be incompatible and yet fit into a consistent theoretical framework. 

\item {\it Identify Linguistic Excess Baggage.}  Remember that human language can contain tacit assumptions that reflect the limited context in which it evolved. We have discussed examples associated with `to happen' and `reality'. Linguistic excess baggage can  be dispensed with in favor of the more precise language of physics.  { \it To clearly  discuss quantum theory, learn to speak the language of  quantum mechanics.} Alternatively linguistic excess baggage can be qualified so that tacit assumptions become explicit. 

\item{\it Identify Dispensible Language.}  Remember that an exposition of a theory can contain language that can be dispensed with without affecting its experimental predictions. Examples are the use of `to happen' to qualify probabilities and the use of `all equally real' to qualify sets of histories. The author often finds it  useful to drop dispensible language in introducing quantum theory. That is especially the case if the audience  carry  their own linguistic baggage which conflicts with the dispensible language making it  an obstacle to understanding and acceptance.
However, don't forget that adding interpretative but dispensible language can be an important route to insight, motivation,  understanding, and generalization. Either way, it is important to recognize when language is dispensible and when it is not. In arguments concerning the interpretation of the theory one then understands which statments are experimentally verifiable and which are not. 

\item{\it Beware of Introspection.} Remember that human beings are physical systems in the universe that have a long, specific, evolutionary history. That evolution is consistent with the universal laws of physics. But the present nature of these IGUSes has far more to do with the frozen accidents of their evolution than with those fundamental laws. As a consequence, introspection is unlikely to be a good guide to their character.  That is especially the case if introspection is seen to require precise representations in the fundamental theory for constructions of the human language. To take an extreme example, some have concluded from the strong impression of `now' held by human IGUSes that the fundamental theory {\it must} incorporate it  despite the overwhelming body of experimental evidence against preferred frames with figures of merit approaching $10^{-21}$ \cite{Will}, and despite the possibility of constructing IGUSes which  do not prefer the present \cite{Har05}. Rather, `now' can be seen as a feature of how certain IGUSes process temporal information as described in Section II. 

\item{\it Beware of Agendas.}  Remember that the basic criteria for physical theory are logical consistency and consistency with experiment. Agendas for physical theory motivate research as in the quest for unification or a selection principle that would distinguish one realm from all others \cite{DK96,Ken96}. But if a theory that is logically consistent and consistent with experiment disagrees with your agenda that  should not be called a problem for the theory, it is more likely to be a problem with your agenda.

\end{itemize}

It seems likely that there are limits to quantum theory and the validity of its principle of superposition as there have been  for every other candidate for a fundamental theory to date. The remarkable fact about the history of  this most successful of all physical theories is that, despite the limited range over which it has been experimentally verified, there are no alternative theories that are consistent these experiments, consistent with the rest of modern physics, but which differ  their predictions in domains not yet tested.  As Steve 
Weinberg puts it: ``It is striking that it has not so far been
possible to find a logically consistent theory that is close to quantum mechanics other than quantum mechanics itself'' \cite{Wei92}. 

Alternatives to quantum theory would be of great interest if only to guide experiment. %The author guesses that these are more likely to arise either from experimental imperatives or from the need to  extend  the present theory to include gravity rather than further intensive analysis of the existing theory. 
Given the  trend in the development of fundamental theory, it is very possible that the disparity between human language and the language of fundamental physics will increase as quantum theory is replaced or extended.  If that is the case, careful analyses of the relationship between human language and the language of  physics of the kind sketched all too superficially in this essay will be increasingly important for clarity of understanding.

\appendix

\renewcommand{\theequation}{\Alph{section}.\arabic{equation}}

\section{The Quantum Mechanics of a Closed System}

This appendix gives a simplified, bare-bones account of some essential elements of the modern synthesis
of ideas constituting the quantum mechanics of closed systems \cite{Gri02, Omn94, Gel94}
using  the model closed box described in Section III.A.

\subsection{Realms}

The most general objective of a quantum theory of a closed system is the prediction of probabilities for the
individual members of exhaustive sets of coarse-grained alternative time histories of the 
system. For instance, we might be interested in alternative histories of the center of mass
of the Earth in its progress around the Sun, or in histories of the correlation between the
registrations of measuring apparatus and a measured subsystem. Alternatives at one moment of
time can always be reduced to a set of yes/no questions.  For example, alternative positions
of the Earth's center of mass can be reduced to asking, ``Is it in this region -- yes or
no?'',
``Is it in that region -- yes or no?'', etc. An exhaustive set of yes/no alternatives is
represented in the Heisenberg picture by an exhaustive set of orthogonal projection operators
$\{P_\alpha(t)\}$,
$\alpha = 1, 2, 3 \cdots$.  These satisfy
\begin{equation}
\sum\nolimits_\alpha P_\alpha(t) = I, \  {\rm and}\ P_\alpha(t)\, P_\beta (t) =
\delta_{\alpha\beta} P_\alpha (t)
\label{aone}
\end{equation}
showing that they represent an exhaustive set of exclusive alternatives.  In the Heisenberg
picture, the operators $P_\alpha(t)$ evolve with time according to
\begin{equation}
P_\alpha(t) = e^{+iHt/\hbar} P_\alpha(0)\, e^{-iHt/\hbar}\, .
\label{atwo}
\end{equation}
The state $|\Psi\rangle$ is unchanging in time.

An important kind of set of histories is specified by a series of sets of 
alternatives $\{P^1_{\alpha_1} (t_1)\}$,
$\{P^2_{\alpha_2} (t_2)\}, \cdots$, $\{P^n_{\alpha_n} (t_n)\}$ at a sequence of times
$t_1<t_2<\cdots < t_n$.  The sets at distinct times can differ, and are distinguished by the
superscript on the $P$'s. For instance, projections on ranges of position might be followed
by projections on ranges of momentum, etc\footnote{In general realistic sets of  histories will be {\it branch dependent} with sets of projections at a given time depending on the particular sequence of previous alternatives, but we ignore this in the present simplified exposition.}. An individual history $c_\alpha$ in such a set is
a
particular sequence of alternatives $(\alpha_1, \alpha_2, \cdots, \alpha_n)\equiv \alpha$
and is represented by the corresponding chain of projections called a {\it class operator}
\begin{equation}
C_\alpha\equiv P^n_{\alpha_n} (t_n) \cdots P_{\alpha_1} (t_1)\, .
\label{athree}
\end{equation}
Such a set of histories is generally {\it coarse-grained} because alternatives are specified
at some times and not at every time, and because the alternatives at a given time are
projections onto subspaces with dimension greater than one and not projections onto a complete
set of states.  {\it Fine-grained} sets of histories consist of one-dimensional
projections at each and every time.

Operations of fine- and coarse-graining may be defined on sets of histories.  
A set of histories $\{c_\alpha\}$ may be {\it coarse-grained} by partitioning it
into exhaustive and exclusive classes $\bar c_{\bar\alpha}, \bar\alpha=1, 2,
\cdots$.  Each class consists of some number of histories in the finer-grained set, and
every finer-grained history is in some class.  Suppose, for example, that the position
of the Earth's center of mass is specified by dividing space into cubical regions of a
certain size. A coarser-grained description of position could consist of
larger regions made up of unions of the smaller ones.  {\it Fine-graining} is the
inverse operation of dividing sets of histories up into smaller classes.  The class
operator $\bar C_{\bar \alpha}$ for a history in a coarse-graining of a set
whose class operators are $\{C_\alpha\}$ is related to those operators by summation,
{\it viz.}
\begin{equation}
\bar C_{\bar\alpha} = \sum_{\alpha\in\bar\alpha} C_\alpha
\label{afour}
\end{equation}
where the sum defining $\bar C_{\bar\alpha}$ for the class $\bar c_{\bar\alpha}$ is the
sum over the $C_\alpha$ for all finer-grained histories contained within it.

For any set of histories $\{c_\alpha\}$, there is a {\it branch state vector} for each
history in the set defined by
\begin{equation}
|\Psi_\alpha\rangle = C_\alpha |\Psi\rangle\, .
\label{afive}
\end{equation}
When probabilities can be consistently assigned to the individual histories in a set,
they are given by
\begin{equation}
p(\alpha) = \parallel |\Psi_\alpha\rangle\parallel^2 =
\parallel C_\alpha |\Psi\rangle\parallel^2\, .
\label{asix}
\end{equation}

However, quantum interference prevents consistent probabilities from being assigned to every set of alternative
histories that may be described.  The two-slit
experiment provides an elementary example: An electron emitted by a source can pass
through either of two slits on its way to detection at a farther screen.  It is not possible to  consistently assign probabilities to the two histories distinguished by which slit the
electron goes through.  Because of interference, 
the probability to arrive at a point
on the screen would not be the sum of the probabilities to arrive there by going
through each of the slits. In quantum theory, probabilities are squares of amplitudes
and the square of a sum is not generally the sum of the squares. On the other hand, if other interactions of the  electron destroy the interference between the two histories (as when a measurement determines which slit it passes through) then probabilities can be consistently assigned.

Negligible interference between the branches of a set
\begin{equation}
\langle\Psi_\alpha|\Psi_\beta\rangle\approx 0 \quad , \quad \alpha\not=\beta
\label{aseven}
\end{equation}
is a sufficient condition for the probabilities \eqref{asix} to be consistent with the
rules of probability theory.
Specifically, as a consequence of the decoherence  condition  \eqref{aseven}, the probabilities \eqref{asix} obey
the most general form of the probability sum rules
\begin{equation}
p(\bar\alpha) \approx \sum_{\alpha\in\bar\alpha} p(\alpha)
\label{aeight}
\end{equation}
for any coarse-graining $\{\bar c_{\bar\alpha}\}$ of the $\{c_\alpha\}$.  Sets of
histories obeying \eqref{aseven} are said to (medium) decohere.\footnote{For a
discussion of the linear positive, weak, medium, and strong decoherence conditions,
see \cite{GH90b, GH95, Har04}. However, as L.~Di\'osi has shown \cite{Dio04}, medium
decoherence is the weakest of this chain that is consistent with elementary notions of the
independence of isolated systems.}
These are sets for which quantum mechanics makes predictions.  They are determined through
\eqref{asix} by the Hamiltonian $H$ and the quantum state of the universe $|\Psi\rangle$.
We use the term {\it realm} as a synonym for a {\it decoherent set of alternative coarse-grained histories}.

A coarse-graining of a decoherent set is again decoherent. A fine-graining of a
decoherent set risks losing decoherence.

An important mechanism of decoherence is the dissipation of phase coherence between
branches into variables not followed by the coarse-graining.  Consider by way of
example, a dust grain in a superposition of two positions deep in interstellar space
\cite{JZ85}.  In our universe, about $10^{11}$ cosmic background photons scatter from
the dust grain each second.  The two positions become correlated with different, nearly
orthogonal states of the photons. Coarse-grainings that follow only the position of the
dust grain at a few times therefore correspond to branch state vectors that are nearly
orthogonal and satisfy \eqref{aseven}.  The orthogonality is approximate but in
realistic situations sufficient to define consistent probabilities well beyond the
standard to which they can be checked or, indeed, the physical situation modeled
\cite{Har91a}.

Measurements and observers play no fundamental role in this general formulation
of  quantum theory. Measurement situations can, of course, be described \cite{GH90a,Har91a}. In a
typical measurement situation, one subsystem of the universe (the measured subsystem)
interacts with another (the apparatus). A variable of the measured subsystem, not otherwise
decohering, becomes correlated with a variable of the apparatus which decoheres because of
its interaction with the rest of the universe. The correlation thus effects the decoherence
of the measured variables so that probabilities can be predicted for its values. With
suitable idealizations and assumptions, probabilities for the measured outcomes are given to
an excellent approximation by usual textbook quantum theory.  But, in a set of histories
where they
decohere, probabilities can be assigned to the position of the Moon when it is not
receiving the attention of observers and to the values of density fluctuations in the early
universe when there were neither measurements taking place nor observers to carry them out.

The probabilities of the histories of the possible {\it decoherent} sets of
coarse-grained histories and the conditional probabilities constructed from them
are the totality of predictions of the quantum mechanics of a closed
system given the Hamiltonian $H$ and initial state $|\Psi\rangle$.

\subsection{Incompatible Realms}

Coarse-graining is generally necessary for decoherence. There are only trivial decoherent
sets of completely fine-grained histories \cite{GH06}. 

A completely fine-grained set of histories can be coarse-grained in many different ways to
yield a decoherent set whose probabilities can be employed in the processes of
prediction and retrodiction. Further, there are many different completely fine-grained sets to start
from corresponding to the different possible choices of one-dimensional projections at each
time arising from different complete sets of commuting observables.  Once
coarse-grained enough to achieve decoherence, further coarse graining
preserves it.  Some decoherent sets can be organized into families connected by the
operations of fine and coarse graining.  Such sets are said to be {\it compatible}. 

Realms for which there is no common finer-grained decoherent sets are
{\it incompatible}.  We may not draw inferences by combining
probabilities from incompatible realms\footnote{As the work of R.~Griffiths
\cite{Gri02} especially has shown, essentially all inconsistencies alleged against consistent
histories quantum mechanics arise from violating this logical prohibition.}.
That would implicitly assume that there are
the probabilities of a finer-grained description which is not available. Incompatible
realms provide different descriptions of the universe.  All of the
totality of incompatible realms are necessary to give a complete account of
the universe because they are, in principle, equally available for exhibiting regularities
and constructing explanations.  Quantum theory does not distinguish one of these realms over another without further criteria.

Incompatibility is not inconsistency in the sense of making different predictions for the
same history. The probability of a history $c_\alpha$ is given by \eqref{asix} in all the realms of which it is a member.

 \subsection{Quasiclassical Realms}

While quantum theory permits a great many incompatible descriptions of a closed system
by different realms, we as human observers utilize mainly realms that are coarse-grainings of one family of compatible sets ---
the quasiclassical realms of everyday experience. These are the sets of decoherent histories
whose probabilities manifest the classical 
regularities of the universe that are exploitable in our various pursuits, in
particular, to get food, reproduce, avoid destruction, and achieve recognition.  These are
the sets of histories defined by quasiclassical  alternatives which our perception is adapted to distinguish. 

Quasiclassical  realms are defined a coarse-graining specifying ranges of values of the variables of classical physics. These include the averages over small volumes of approximately conserved quantities such as energy, momentum, and various kind of particle species. We call such variables {\it quasiclassical variables}. The useful properties of  quasiclassical realms follow from the approximate conservation of the variables that define them.
In particular quasiclassical realms exhibit correlations in time governed by the approximate deterministic laws of motion of classical physics.  In our quantum universe, classical laws are applicable over a wide range of time,place, scale, and epoch. 

More specifically,  by a {\it quasiclassical realm} we mean an exhaustive set of mutually exclusive
coarse-grained alternative histories, that obey a realistic principle of decoherence, that
consist largely of related but branch-dependent projections onto ranges  of quasiclassical variables at a succession of times, with
individual histories exhibiting patterns of correlation implied by closed sets of effective
equations of motion subject to frequent small fluctuations and occasional major branchings  (as in measurement situations). By a {\it family of quasiclassical realms} we mean a set of compatible ones that  are all  coarse grainings of a common one. 
  Thus defined, the quasiclassical
realms are a feature of our universe that arise from its
quantum state and dynamics which we are  adapted to exploit \cite{GH06}.

\acknowledgments
The author is grateful to Fay Dowker, Simon Saunders, and  Rafael Sorkin for critical readings of the manuscript. 
Thanks are due to David Deutsch for clarifying discussions of the many worlds formulation of quantum theory and to Simon Saunders help with the references. The author thanks the Aspen Center for Physics for hospitality while this work was being completed. 
This work was supported in part by the National Science Foundation under grants
PHY02-44764 and PHY05-55669.

\end{document}